\documentclass[12pt]{amsart}

\usepackage{amsmath}
\usepackage{amssymb}
\usepackage{amsthm}
\usepackage{enumerate}
\usepackage{graphicx}
\usepackage{tikz-cd}
\usepackage{bbm}
\usepackage{xcolor}
\usepackage{setspace}
\usepackage{chngcntr}
\usepackage{float}
\usepackage{hyperref}
\hypersetup{colorlinks=true, linkcolor=blue}

\newtheorem{theorem}{Theorem}
\newtheorem*{theorem*}{Theorem}
\newtheorem*{lemma*}{Lemma}
\newtheorem*{prop*}{Proposition}
\newtheorem*{cor*}{Corollary}
\newtheorem*{claim*}{Claim}

\newtheorem{conjecture}[theorem]{Conjecture}

\newtheorem{lemma}[theorem]{Lemma}

\theoremstyle{definition}
\newtheorem{definition}{Definition}

\newtheorem*{definition*}{Definition}
\newtheorem*{remark*}{Remark}
\newtheorem*{hint*}{Hint}
\newtheorem*{warning*}{Warning}
\newtheorem*{example*}{Example}

\definecolor{darkgreen}{rgb}{.2,.7,.2}

\allowdisplaybreaks

\def\sm{\setminus}

\def\NN{{\mathbb N}}

\def\RR{{\mathbb R}}
\def\SS{{\mathbb S}}

\def\1{{\mathbbm 1}}

\def\cal{\mathcal}

\def\ol{\overline}

\def\tr{\operatorname{tr}}

\def\div{\operatorname{div}}
\def\Rc{\operatorname{Rc}} %Ricci curvature

\def\wt{\widetilde}

\def\wtK{{\widetilde{K}}}

\def\wtM{{\widetilde{M}}}

\def\wh{\widehat}

\def\whM{{\widehat{M}}}

\title{Causal geodesic incompleteness of spacetimes arising from IMP gluing}
\author{Madeleine Burkhart}
\address[Madeleine Burkhart]{University of Washington, Seattle, USA}
\author{Daniel Pollack}
\address[Daniel Pollack]{University of Washington, Seattle, USA}

\begin{document}

\maketitle

\begin{abstract}
In 2002, Isenberg-Mazzeo-Pollack (IMP) constructed a series of vacuum initial data sets via a gluing construction. In this paper, we investigate some local geometry of these initial data sets as well as implications regarding their spacetime developments. In particular, we state conditions for the existence of outer trapped surfaces near the center of the IMP gluing neck and thence use a generalization of the Penrose incompleteness theorem to deduce null incompleteness of the resulting spacetimes.
\end{abstract}

\section*{Acknowledgements}
This work was supported by a grant from the Simons Foundation (279720-DP). Both authors would like to thank Greg Galloway for his many helpful suggestions.

\section{Introduction} \label{sec:intro}

A natural question in mathematical general relativity concerning the construction of initial data sets (IDSs) is the following: Given two IDSs $(M_{1},\gamma_{1},K_{1})$ and $(M_{2},\gamma_{2},K_{2})$ satisfying compatible forms of the Einstein constraint equations (ECE), is it possible to glue the data together to form an IDS with the topology $M_{1}\#M_{2}$?

This question has been investigated in the work of Chru{\'s}ciel, Corvino, Delay, Eichmair, Isenberg, Maxwell, Mazzeo, Miao, Pollack, and Schoen (see \cite{CD03}, \cite{CIP05}, \cite{C00}, \cite{CEM13}, \cite{CS06}, \cite{IMaxP05}, \cite{IMP02}, and \cite{IMP03}). Notably, the work of Chru{\'s}ciel-Isenberg-Pollack (CIP) \cite{CIP05}---which builds upon the basic structure of the Isenberg-Mazzeo-Pollack \cite{IMP02} construction and the Corvino \cite{C00} and Chru{\'s}ciel-Delay \cite{CD03} localizing procedure---was used to find examples of spacetimes that do not contain constant mean curvature (CMC) Cauchy surfaces.

We will focus on gluing constructions for vacuum IDSs. That is, we require that $(M_{i},\gamma_{i},K_{i})$ satisfies the vacuum constraints:
\begin{align}
\div K_{i}-\nabla\tr K_{i}&=0, \label{aln:vmom} \\
R_{\gamma_{i}}-|K_{i}|_{\gamma_{i}}^{2}+(\tr K_{i})^{2}&=0. \label{aln:ven}
\end{align}

Given the above constructions, a subsequent question we can ask is what we might expect from spacetimes that evolve from such glued IDSs. In particular, gluing constructions tend to involve the formation of a neck between $M_{1}$ and $M_{2}$ with significantly warped geometry; heuristically, we may expect spacetimes arising from these IDSs to be causally incomplete.

This secondary question has been of interest lately due to the challenging problem of existence and genericity of CMC Cauchy surfaces. In particular, a recent result of Galloway-Ling '18 \cite{GL18(1)} establishes sufficient conditions for existence of CMC slices in cosmological spacetimes; one such condition is future timelike geodesic completeness. Galloway and Ling also state a conjecture relating the two concepts:
\begin{conjecture}[Galloway-Ling] \label{conj:GL}
Let $(\whM,g)$ be a spacetime with compact Cauchy surfaces. If $(\whM,g)$ is future timelike geodesically complete and satisfies the strong energy condition, i.e., $\wh{\Rc}(X,X)\ge0$ for all timelike $X$, then $(\whM,g)$ contains a CMC Cauchy surface.
\end{conjecture}

If this conjecture is proven, it would both generalize their existence result and imply the Bartnik splitting conjecture \cite{B882}.

Now, the two main sources of counterexamples to the existence of CMC Cauchy surfaces are Bartnik \cite{B882} and CIP \cite{CIP05}. The former examples of spacetimes are constructed to be timelike geodesically incomplete, while the work of Burkhart-Lesourd-Pollack '19 \cite{BLP19} establishes null geodesic incompleteness of the latter, given sufficiently large gluing parameter.

This most recent incompleteness result is done by exploiting symmetry of the specific CIP examples, but, as stated above, one expects that all spacetimes arising from IMP gluing should be incomplete due to the pinched geometry of the gluing neck. The main result of this paper is to prove that such spacetimes are indeed null geodesically incomplete:

\begin{theorem} \label{thm:BP}
The compact vacuum IDSs $(\wtM^{n},\wt{\gamma}_{T},\wtK_{T})$ in \cite{IMP02} and \cite{CIP05} contain surfaces, diffeomorphic to $\SS^{n-1}$, near the center of the IMP gluing neck, which are trapped with respect to the unit null normal pointing toward the center of the neck. Thus, we may conclude the following:
\begin{enumerate}[(1)]
\item If $\wtM$ is formed by gluing disjoint IDSs, and if $\wtM$ has a noncompact cover, any spacetime development of the IDS is both future and past null geodesically incomplete.
\item For $3\le n\le7$, there exists a stable MOTS, diffeomorphic to $\SS^{n-1}$, near the center of the IMP gluing neck in $\wtM$.
\end{enumerate}
\end{theorem}

While the Burkhart-Lesourd-Pollack result utilizes symmetry of the CIP construction, the above result comes down to estimating the null expansions of certain cross-sections in the IMP gluing neck and concluding that these cross-sections must be trapped in the null direction pointing toward the center of the neck. Using a generalization of the Penrose incompleteness theorem which can be found in \cite{G14}, we conclude future null incompleteness of spacetime developments of these gluing constructions given a noncompactness condition. In addition, the MOTS existence result of Andersson-Eichmair-Metzger \cite{AEM11} thence implies the existence of a stable MOTS near the center of the gluing neck in dimensions $3\le n\le 7$.

We make two remarks on generalizing Theorem \ref{thm:BP}: First, while we focus on the compact IDSs from \cite{IMP02}, the analogous result will also hold for asymptotically Euclidean and hyperbolic IDSs, since the same estimates hold on the analogous H{\"o}lder spaces in those cases. Second, we have only shown the result for vacuum IDSs, but it would also be desirable to prove a similar result for the non-vacuum glued IDSs constructed by Iseberg-Maxwell-Pollack \cite{IMaxP05}. Beyond this, in the interest of resolving Conjecture \ref{conj:GL}, it would be nice to resolve timelike incompleteness of the CIP examples, though this is much more difficult from an initial data perspective.

In section \ref{sec:back}, we set up notation and describe the IMP gluing procedure as well as the incompleteness result we use. Then, in section \ref{sec:results}, we prove Theorem \ref{thm:BP} by computing the null expansions of particular cross-sections. Throughout, while we primarily focus on IMP gluing, we also indicate how the argument is altered for the CIP construction.

\section{Preliminaries} \label{sec:back}

First off, since we consider $(n-1)$, $n$, and $(n+1)$-dimensional objects, we use the following index convention: Greek indices run from 0 to $n$, Latin indices near the middle of the alphabet (specifically, $i$ to $t$) run from 1 to $n$, and Latin indices near the front and back ends of the alphabet ($a$ to $h$ and $u$ to $z$) run from 1 to $n-1$. As for distinguishing between objects in different dimensions, we generally use hats to refer to objects on a Lorentzian manifold and subscripts when restricting a tensor on an initial data set $(M^{n},\gamma,K)$ to an $(n-1)$-dimensional submanifold.

The IMP gluing construction uses the conformal method to obtain vacuum data on connected sums, so before jumping into the former, we recall the conformal method for solving the vacuum constraint equations: given a Riemannian manifold $(M^{n},\gamma)$ with data $K=\mu+{\tau\over n}\gamma$, where $\mu$ is a transverse-traceless (that is, traceless and divergence-free) symmetric covariant 2-tensor and $\tau$ is a function, consider the data $(\psi,X)$, where $\psi$ is a positive function and $X$ is a vector field. Now set:
\begin{align}
\wt{\gamma}&=\psi^{q}\gamma,\quad\text{and} \label{aln:confmetric} \\
\wtK&=\psi^{-2}(\mu+\cal{D}X)+{\tau\over n}\psi^{q}\gamma, \label{aln:conf2ff}
\end{align}
where $q={4\over n-2}$ and $\cal{D}$ is the conformal Killing operator, given by
\begin{equation} \label{eqn:confKill}
(\cal{D}W)_{ij}=X_{i;j}+X_{j;i}-{2\over n}\div_{\gamma}(X)\gamma_{ij}.
\end{equation}
Then if $(\psi,X)$ solve the following equations:
\begin{align}
&\div_{\gamma}(\cal{D}X)={n-1\over n}\psi^{q+2}\nabla\tau, \quad\text{and} \label{aln:divDX} \\
&\Delta_{\gamma}\psi-{1\over q(n-1)}R_{\gamma}\psi+{1\over q(n-1)}|\mu+\cal{D}X|_{\gamma}^{2}\psi^{-q-3}-{1\over qn}\tau^{2}\psi^{q+1}=0, \label{aln:Lich}
\end{align}
the initial data set given by $(M,\wt{\gamma},\wtK)$ satisfies the vacuum constraint equations (\ref{aln:vmom}) and (\ref{aln:ven}). When $\tau$ is constant, observe that we may take $X\equiv0$, so Equation (\ref{aln:divDX}) completely disappears. The resulting version of (\ref{aln:Lich}) is called the Lichnerowicz equation. This simplification is a major reason for the active research area of understanding when spacetimes admit CMC Cauchy surfaces.

\subsection{IMP gluing} \label{subsec:gluing}

Here we recall the gluing construction of \cite{IMP02}, noting the departure points of subsequent generalizations. We start with an IDS $(M^{n},\gamma,K)$ which satisfies the vacuum Einstein constraint equations, where $M$ is compact and may or may not be connected: when $M$ is connected, this gluing construction creates a `wormhole' in the IDS. We remark that although IMP gluing can be done for $M$ connected, our result requires that $M=M_{1}\coprod M_{2}$ consists of two disjoint components, so as to satisfy the `separating' condition of the generalization of Penrose, which we state in Subsection \ref{subsec:trap}.

Decompose $K$ as follows:
\begin{equation*}
K=\mu+{\tau\over n}\gamma,
\end{equation*}
where $\mu$ is transverse-traceless and $\tau=\tr_{\gamma}K$ is the mean curvature of $M$ in any spacetime development: in this version of IMP gluing, we assume $\tau$ is constant: in \cite{IMP03}, the construction is generalized to require only that $\tau$ be constant around the gluing region, with no adjustments in the gluing procedure or estimates, while \cite{CIP05} has no assumptions on $\tau$ being constant: instead, that construction first uses \cite{B881} to perturb neighborhoods around the gluing region of the initial data in any spacetime evolution, obtaining constant $\tau$ in these regions.

Before diving into the construction, we place one additional restriction upon $(M,\gamma,K)$: in \cite{IMP02}, there is a nondegeneracy condition imposed (Definition 1 in \cite{IMP02}) as well as an assumption that $K\not\equiv0$. In \cite{CIP05}, the nondegeneracy condition is weakened to a no-KIDs requirement in the gluing region: That is, letting $P$ be the linearization of the map taking $(\gamma,K)$ to the constraints $(\rho,J)$ and letting $P^{*}$ be its formal adjoint, we require that the set of Killing Initial Data (KID), or solutions $(N,Y)$ to the equations $P^{*}(N,Y)=0$, be trivial in the gluing region.

\subsubsection{The gluing process} \label{subsubsec:gluing}

We now set up the geometry of the gluing construction: Two points, $p_{1}\in M_{1}$ and $p_{2}\in M_{2}$, are distinguished, and disjoint normal neighborhoods of radius $2R$ are taken around these points. Let $r_{1}$ and $r_{2}$ be the corresponding radial parameters on these normal neighborhoods, so that $\gamma$ restricted to these neighborhoods is
\begin{equation} \label{eqn:normaldecomp}
\gamma|_{B_{2R}(p_{j})}=dr_{j}^{2}+r_{j}^{2}h_{j}(r_{j}),
\end{equation}
where $h_{j}:[0,2R]\to\Gamma(\Sigma^{2}TM)$ is a smooth family of metrics on $\SS^{n-1}$ such that $h_{j}(0)$ is the standard round metric.

First we transform these normal neighborhoods (minus a point) to asymptotically cylindrical tubes. Consider the conformal factor
\begin{equation} \label{eqn:psic}
\psi_{c}(p):=\begin{cases}
1 & p\in M\sm(B_{2R}(p_{1})\cup B_{2R}(p_{2})) \\
\text{interpolation} & p\in B_{2R}(p_{j})\sm B_{R}(p_{j}) \\
r_{j}^{2/q} & p\in B_{R}(p_{j}),
\end{cases}
\end{equation}
where $q={4\over n-2}$ as above, and by interpolation, we here and henceforth mean interpolation of the explicitly defined functions using radial cutoff functions with bounded derivatives. Now in preparation for the gluing, we let
\begin{equation} \label{eqn:blownup}
\gamma_{c}=\psi_{c}^{-q}\gamma\quad\text{and}\quad K_{c}=\psi_{c}^{2}\mu+{\tau\over n}\psi_{c}^{-q}\gamma
\end{equation}
on $M\sm\{p_{1},p_{2}\}$. This choice of conformal factor is chosen so that the resulting regions $(B_{R}(p_{j})\sm\{p_{j}\},\gamma_{c})$ are asymptotically cylindrical for $r_{j}\searrow0$.

Before cutting off the cylinders and applying the quotient map to glue, we introduce the parameter $t_{j}=-\log r_{j}$, so in terms of $t_{j}$, we have the metric decomposition
\begin{equation} \label{eqn:cdecomp}
\gamma_{c}|_{B_{R}(p_{j})\sm\{p_{j}\}}=dt_{j}^{2}+h_{j}(e^{-t_{j}}),
\end{equation}
and the following relation holds: $\partial_{t_{j}}=-r_{j}\partial_{r_{j}}$. We now cut each asymptotically cylindrical end off at $t_{j}=T-\log(R)$. We note that in the subsequent analysis, including the results of this paper, $T$ is a parameter that is eventually taken to be sufficiently large. In accord, with \cite{IMP02}, we denote $T$ as the \textit{gluing parameter}. We now topologically glue by identifying
\begin{equation*}
(t_{1},\theta_{1})\sim(t_{2},\theta_{2})=(T-2\log(R)-t_{1},-\theta_{1}).
\end{equation*}
Note that the identification $\theta_{2}=-\theta_{1}$ is due to a reversal of orientation when gluing. At this juncture we introduce a new parameter $s$ given by
\begin{equation*}
s=t_{1}+\log(R)-T/2=T/2-\log(R)-t_{2},
\end{equation*}
so that $s=0$ corresponds to the center of the glued tube and $s=-T/2,T/2$ correspond to $r_{1},r_{2}=R$, respectively. Explicitly, we have
\begin{equation} \label{eqn:sr}
r_{1}={R\over e^{T/2}}e^{-s}\quad\text{and}\quad r_{2}={R\over e^{T/2}}e^{s},\quad\text{which yields}\quad r_{1}={R^{2}\over e^{T} r_{2}}.
\end{equation}
Denote the glued manifold as $\wtM$ and the tube region given by the image of $B_{R}(p_{1})\cup B_{R}(p_{2})\sm\{p_{1},p_{2}\}$ by $C_{T}$. Also denote the image of $M\sm(B_{r}(p_{1})\cup B_{r}(p_{2}))$ by $M_{r}^{*}$ and define $Q_{a}:=(a-1,a+1)\times\SS^{n}\subset C_{T}$.
The new metric and second fundamental form data on $\wtM$ are given by
\begin{equation} \label{eqn:approxsolns}
\gamma_{T}=\chi_{1}\gamma_{1}+\chi_{2}\gamma_{2}\quad\text{and}\quad\mu_{T}=\chi_{1}\mu_{1}+\chi_{2}\mu_{2},
\end{equation}
where $\gamma_{j}=\gamma_{c}|_{B_{R}(p_{j})}$, $\mu_{j}=\psi_{c}^{2}\mu|_{B_{R}(p_{j})}$, and $\{\chi_{1},\chi_{2}\}$ is a partition of unity with respect to an open cover of $\wtM$ whose intersection consists of $Q_{0}$ and a set disjoint from $C_{T}$.

In addition to this data, a new conformal factor is defined using the cutoff functions $\{\wt{\chi}_{1},\wt{\chi}_{2}\}$, such that $\wt{\chi}_{1}+\wt{\chi}_{2}=1$ on $M_{R}^{*}$, but on $C_{T}$,
\begin{align*}
\wt{\chi}_{1}&=\begin{cases}
1 & s\in[-T/2,T/2-1) \\
\text{interpolation} & s\in[T/2-1,T/2) \\
0 & s=T/2
\end{cases}\quad\text{and} \\
\wt{\chi}_{2}&=\begin{cases}
0 & s=-T/2 \\
\text{interpolation} & s\in(-T/2,1-T/2] \\
1 & s\in(1-T/2,T/2]
\end{cases}.
\end{align*}
We remark that the above are cutoff functions, but seeing as they add to 2 in most of $C_{T}$, they are decidedly not a partition of unity. We now define an approximate conformal factor by
\begin{equation*}
\psi_{T}=\wt{\chi}_{1}\psi_{1}+\wt{\chi}_{2}\psi_{2},
\end{equation*}
where $\psi_{j}=\psi_{c}|_{B_{R}(p_{j})}$. Unwinding this definition and considering when the cutoff functions are $\equiv1$, we see that
\begin{equation*}
\psi_{T}=\begin{cases}
r_{1}^{2/q}+\wt{\chi}_{2}r_{2}^{2/q} & s\in[-T/2,1-T/2) \\
r_{1}^{2/q}+r_{2}^{2/q} & s\in[1-T/2,T/2-1] \\
\wt{\chi}_{1}r_{1}^{2/q}+r_{2}^{2/q} & s\in(T/2-1,T/2]
\end{cases}.
\end{equation*}
Splitting the gluing region into the sections $s\in[-T/2,-1]$ and $s\in[1,T/2]$ and using (\ref{eqn:sr}), we obtain
\begin{equation} \label{eqn:psir1}
\psi_{T}=\begin{cases}
r_{1}^{2/q}+\wt{\chi}_{2}\left({R^{2}\over e^{T}r_{1}}\right)^{2/q} & s\in[-T/2,1-T/2) \quad \left(r_{1}\in\left[{R\over e},R\right]\right) \\
r_{1}^{2/q}+\left({R^{2}\over e^{T}r_{1}}\right)^{2/q} & s\in[1-T/2,-1] \quad \left(r_{1}\in\left[{R\over e^{T/2-1}},{R\over e}\right]\right)
\end{cases}
\end{equation}
and
\begin{equation} \label{eqn:psir2}
\psi_{T}=\begin{cases}
\left({R^{2}\over e^{T}r_{2}}\right)^{2/q}+r_{2}^{2/q} & s\in[1,T/2-1] \quad \left(r_{2}\in\left[{R\over e^{T/2-1}},{R\over e}\right]\right) \\
\wt{\chi}_{1}\left({R^{2}\over e^{T}r_{2}}\right)^{2/q}+r_{2}^{2/q} & s\in[T/2-1,T/2] \quad \left(r_{2}\in\left[{R\over e},R\right]\right)
\end{cases}.
\end{equation}

\subsubsection{Perturbing to a solution} \label{subsubsec:perturb}

Now that all the data has been patched, we can consider an approximate solution given by:
\begin{equation*}
\left(\wtM,\psi_{T}^{q}\gamma_{T},\psi_{T}^{-2}\mu_{T}+{\tau\over n}\gamma_{T}\right).
\end{equation*}
In order to turn this into an exact solution to the vacuum Einstein constraint equations, we use the conformal method. That is, we must perturb $\mu_{T}$ and $\psi_{T}$ so that the former is transverse-traceless and the latter solves the Lichnerowicz equation. Arguments as to why such perturbations exist for large enough gluing parameter form the bulk of the foundational IMP paper \cite{IMP02}. We remark that the approximate data still satisfies the constraints on $M_{2R}^{*}$, that $\mu_{T}$ is still transverse-traceless on $\wtM\sm Q_{0}$, and that $\psi_{T}$ satisfies the Lichnerowicz equation on $M_{R}^{*}$, which, using \cite{CD03} and \cite{CIP05} to localize the global IMP arguments, is why we only need the CMC and no-KID conditions mentioned above to hold around the gluing region for all the analysis to go through in the CIP construction.

We denote the perturbations on $\mu_{T}$ and $\psi_{T}$ by $\sigma_{T}$ and $\eta_{T}$, respectively. In our subsequent computations on the glued data, we will need H{\"o}lder estimates on these perturbation terms. To this end, we make the following definitions (from \cite{IMP02}):

\begin{definition} \label{def:Holder}
Let $||X||_{k,\alpha,\Omega}$ denote the H{\"o}lder norm with respect to $\gamma_{T}$ of a vector field $X$ on an open subset $\Omega\subset\wtM$ for $k\in\NN$ and $\alpha\in(0,1)$. We define
\begin{equation*}
||X||_{k,\alpha}:=||X|_{M_{R/2}^{*}}||_{k,\alpha}+\sup\limits_{1-T/2\le a\le T/2-1}||X|_{C_{T}}||_{k,\alpha,Q_{a}}.
\end{equation*}
The corresponding H{\"o}lder norm on tensors is induced from the above definition.
\end{definition}

The above norm is useful because it gives us some control over objects on the gluing neck. We define a weighted H{\"o}lder norm that similarly takes growth along the neck into account.

\begin{definition} \label{def:wgtedHolder}
Using the terminology of Definition \ref{def:Holder}, define a weighting function by
\begin{equation*}
w_{T}(p):=\begin{cases}
e^{-T/q}\cosh(2s/q) & p=(s,\theta)\in C_{T} \\
\text{interpolation} & p\in M_{R}^{*}\sm M_{2R}^{*} \\
1 & p\in M_{2R}^{*}
\end{cases}.
\end{equation*}
For any $k\in\NN$, $\alpha\in(0,1)$, $\delta\in\RR$, and $\phi\in C^{k,\alpha}(M_{T})$, define
\begin{equation*}
||\phi||_{k,\alpha,\delta}:=||w_{T}^{-\delta}\phi||_{k,\alpha}.
\end{equation*}
\end{definition}

When making estimates, we often use the expression ``$X\lesssim Y$," which means ``$X$ is less than a positive constant times $Y$." When it is unclear, we will specify what parameters that constant may depend on, and in some of the more arduous computations that follow, we will eschew this notation and write $X\le C(\cdot,\ldots,\cdot)Y$. Because the order estimates of Section \ref{sec:results} will depend primarily upon the parameter $T$, the most important thing to emphasize is that none of these constants depend on $T$.

Back to the perturbations themselves, \cite{IMP02} and \cite{IMaxP05} obtain the covariant 2-tensor $\sigma_{T}$ so that
\begin{equation} \label{eqn:sigma}
\wt{\mu}_{T}:=\mu_{T}-\sigma_{T}
\end{equation}
is transverse traceless. This is done by solving the elliptic equation
\begin{equation*}
LX:=\cal{D}^{*}\circ\cal{D}X=W,
\end{equation*}
where $\cal{D}$ is the conformal Killing operator as in (\ref{eqn:confKill}), $\cal{D}^{*}:=-\div_{\gamma_{T}}$ is its formal adjoint, and $W:=(\div_{\gamma_{T}}\mu_{T})^{\#}$. This yields the desired perturbation $\sigma_{T}:=\cal{D}X$. In \cite{CIP05}, the perturbation is localized by considering the corresponding boundary value problem to ensure $\sigma_{T}$ is 0 on the boundary of the gluing region, and then using the smoothing procedure of \cite{CD03} across boundary spheres. We note that the smoothing is compactly supported away from the central part of the gluing neck, so it does not affect the areas where we make estimates in Section \ref{sec:results}.

Before moving onto the conformal factor, we state bounds on the perturbation $\sigma_{T}$ which we will use in Section \ref{sec:results}. These are generally contained and discussed in \cite{IMaxP05}, but we include full justification for the sake of completeness. From Lemma 3.2 in \cite{IMaxP05} and following the subsequent remarks, we see that $\sigma_{T}$ satisfies the following H{\"o}lder bound:
\begin{align} \label{eqn:sigmaHolderbd}
||\sigma_{T}||_{k,\alpha}&=||\cal{D}X||_{k,\alpha} \\
&\lesssim||X||_{k+1,\alpha} \nonumber \\
&\lesssim T^{3}||W||_{k-1,\alpha} \nonumber \\
&=T^{3}||(\div\mu_{T})^{\#}||_{k-1,\alpha} \nonumber \\
&\lesssim T^{3}||\mu_{T}||_{k,\alpha,Q_{0}}. \nonumber
\end{align}
where all constants are independent of $T$ and the final inequality follows because $\mu_{T}$ is transverse traceless away from $Q_{0}$. Now we work on estimating $||\mu_{T}||_{k,\alpha,Q_{0}}$. Using (\ref{eqn:approxsolns}), (\ref{eqn:blownup}), (\ref{eqn:psic}), and (\ref{eqn:sr}), we obtain
\begin{align*}
\mu_{T}&=\chi_{1}\mu_{1}+\chi_{2}\mu_{2} \\
&=\chi_{1}\psi_{c}^{2}\mu|_{B_{R}(p_{1})}+\chi_{2}\psi_{c}^{2}\mu|_{B_{R}(p_{2})} \\
&=\chi_{1}r_{1}^{n-2}\mu|_{B_{R}(p_{1})}+\chi_{2}r_{2}^{n-2}\mu|_{B_{R}(p_{2})} \\
&=\chi_{1}\left({R\over e^{T/2}}e^{-s}\right)^{n-2}\mu|_{B_{R}(p_{1})}+\chi_{2}\left({R\over e^{T/2}}e^{s}\right)^{n-2}\mu|_{B_{R}(p_{2})} \\
&=\left(\chi_{1}(Re^{-s})^{n-2}\mu|_{B_{R}(p_{1})}+\chi_{2}(Re^{s})^{n-2}\mu|_{B_{R}(p_{2})}\right)e^{-{T(n-2)\over2}},
\end{align*}
and
\begin{align*}
\gamma_{T}&=\chi_{1}\gamma_{1}+\chi_{2}\gamma_{2} \\
&=\chi_{1}\psi_{c}^{-q}\gamma|_{B_{R}(p_{1})}+\chi_{2}\psi_{c}^{-q}\gamma|_{B_{R}(p_{2})} \\
&=\chi_{1}r_{1}^{-2}\gamma|_{B_{R}(p_{1})}+\chi_{2}r_{2}^{-2}\gamma|_{B_{R}(p_{2})} \\
&=\chi_{1}\left({R\over e^{T/2}}e^{-s}\right)^{-2}\gamma|_{B_{R}(p_{1})}+\chi_{2}\left({R\over e^{T/2}}e^{s}\right)^{-2}\gamma|_{B_{R}(p_{2})} \\
&=\left(\chi_{1}(Re^{-s})^{-2}\gamma|_{B_{R}(p_{1})}+\chi_{2}(Re^{s})^{-2}\gamma|_{B_{R}(p_{2})}\right)e^{T},
\end{align*}
so on $Q_{0}$, this yields $\mu_{T}=e^{-nT/2+T}\zeta$ and $\gamma_{T}^{-1}=e^{-T}\xi$, where $\zeta$ and $\xi$ are smooth with $0^{th}$ through $(k+1)^{st}$ derivatives uniformly bounded by constants independent of $T$. Thus, from (\ref{eqn:sigmaHolderbd}) we have the following bound on $\sigma_{T}$:
\begin{equation} \label{eqn:sigmabd}
||\sigma_{T}||_{k,\alpha}\lesssim T^{3}||\mu_{T}||_{k,\alpha,Q_{0}}\lesssim T^{3}e^{-nT/2}.
\end{equation}

We now turn our attention to the perturbed conformal factor
\begin{equation} \label{eqn:eta}
\wt{\psi}_{T}:=\psi_{T}+\eta_{T},
\end{equation}
which is obtained by solving the Lichnerowicz equation for the data $(\wtM,\gamma_{T},\wt{\mu}_{T}+{\tau\over n}\gamma_{T})$. Again, in \cite{CIP05}, the perturbation is localized by considering the corresponding boundary value problem and smoothing procedure. From section 3.4.3 of \cite{IMaxP05}, we see that the perturbation $\eta_{T}$ satisfies the bound
\begin{equation} \label{eqn:etabds}
||\eta_{T}||_{k,\alpha,\delta}:=||w_{T}^{-\delta}\eta_{T}||_{k,\alpha}\lesssim e^{(-\lambda+\delta/q)T},
\end{equation}
where $\lambda>1/q$.

Since $\wt{\mu}_{T}$ is transverse-traceless and $\wt{\psi}_{T}$ satisfies the Lichnerowicz equation, we have by the conformal method that for $T$ sufficiently large, the initial data set $(\wtM,\wt{\gamma}_{T},\wt{K}_{T})$ given by
\begin{equation} \label{eqn:finalmetric}
\wt{\gamma}_{T}=\wt{\psi}_{T}^{q}\gamma_{T}\quad\text{and}\quad \wt{K}_{T}=\wt{\psi}_{T}^{-2}\wt{\mu}_{T}+{\tau\over n}\wt{\gamma}_{T}
\end{equation}
satisfies the vacuum Einstein constraint equations. This is the glued initial data set we investigate in Section \ref{sec:results}.

\subsection{Outer/inner trapped surfaces} \label{subsec:trap}

The main result of this paper involves finding outer and inner (depending on the choice of normal) trapped surfaces in the gluing necks of IDSs constructed as above. Before proceeding to prove the result, we review notation having to do with the theory of trapped surfaces.

Given a closed hypersurface $\Sigma^{n-1}$ with trivial normal bundle in an IDS $(M^{n},\gamma,K)$, recall that the outer null expansion of $\Sigma$ is given by
\begin{equation*}
\theta=\theta^{+}=\tr_{\Sigma}K+H_{\Sigma},
\end{equation*}
where $H_{\Sigma}$ is the mean curvature of $\Sigma$ in $M$ with respect to the outward-pointing unit normal. We may also regard the inner null expansion, given by $\theta^{-}=\tr_{\Sigma}K-H_{\Sigma}$. Heuristically, if we assume $M$ is embedded in a spacetime, the null expansions measure how the area of $\Sigma$ changes along outward and inward-pointing light rays. As such, we make the following standard definitions:

\begin{definition}
If $\theta^{+}<0$, we call $\Sigma$ \textit{outer trapped} and if $\theta^{-}<0$, $\Sigma$ is \textit{inner trapped}. If both $\theta^{+}<0$ and $\theta^{-}<0$, we call $\Sigma$ \textit{trapped}. If $\theta^{+}=0$, we say $\Sigma$ is a \textit{marginally outer trapped surface (MOTS)}, and likewise if $\theta^{-}=0$, $\Sigma$ is a \textit{marginally inner trapped surface (MITS)}.
\end{definition}

We will use the following generalization of the Penrose incompleteness theorem, which can be found in \cite{G14}:

\begin{theorem} (Chru{\'s}ciel-Galloway) \label{thm:outerpenrose}
Let $(\whM,g)$ be a globally hyperbolic spacetime and suppose the following hold:
\begin{enumerate}[(i)]
\item $\whM$ admits a non-compact Cauchy surface $M$.
\item $\whM$ obeys the null energy condition.
\item $M$ contains a closed, connected hypersurface $\Sigma$ that is \textit{separating}---that is, $M\sm\Sigma=U\cup V$, where $U$ and $V$ are disjoint connected open sets and $U$ has non-compact closure---and $\Sigma$ is outer trapped with respect to the null normal $\ell^{+}$ pointing toward $U$.
\end{enumerate}
Then $(\whM,g)$ is future null geodesically incomplete.
\end{theorem}

We use the above to obtain incompleteness in the general gluing construction, since we are only able to find surfaces that are outer trapped. Notice that although we focus on the gluing construction in the case of compact Cauchy surfaces, the above theorem requires non-compactness: to get around this we pass to a non-compact cover when it exists. This is why we require that the IDSs we investigate admit non-compact covers.

\section{Future null incompleteness of gluing constructions} \label{sec:results}

In this section, we prove Theorem \ref{thm:BP}. To do this, we first show in Subsection \ref{subsec:showingtrapped} that the IMP gluing construction always yields cross sections (corresponding to $s=\pm1$) near the middle of the gluing neck that are inner and outer trapped, respectively. Then, in Subsection \ref{subsec:incompleteness}, we use Theorem \ref{thm:outerpenrose} to deduce null incompleteness when $\wtM$ has a noncompact cover. Lastly, using the MOTS existence result of Anderson-Eichmair-Metzger \cite{AEM11}, we conclude in Subsection \ref{subsec:MOTSexist} that in dimensions $3\le n\le7$, there exists a stable MOTS in the central region $Q_{0}$ of the gluing neck.

\subsection{Obtaining outer and inner trapped surfaces} \label{subsec:showingtrapped}

In order to obtain outer and inner trapped surfaces, we estimate the null expansions at the parameters $s=\pm1$ and show that these cross-sections are trapped in the direction pointing to the center (ie, $s=0$) of the gluing neck.

The first two steps (\ref{subsubsec:trKcomp} and \ref{subsubsec:Hcomp}) are to compute $\tr_{\wt{\gamma}_{T}|_{s=s_{0}}}\wtK_{T}$ and $H_{\wt{\gamma}_{T}|_{s=s_{0}}}$ at arbitrary cross-sections $s=s_{0}$, where $s_{0}\in[-T/2,-1]\cup[1,T/2]$. Then in \ref{subsubsec:estimate} we estimate the results for $s_{0}=\pm1$ with bounds in terms of $T$.

\subsubsection{Computing $\tr_{\wt{\gamma}_{T}|_{s=s_{0}}}\wtK_{T}$} \label{subsubsec:trKcomp}

Unwinding the definitions given by Equations (\ref{eqn:sigma}), (\ref{eqn:eta}), and (\ref{eqn:finalmetric}), we have

\begin{equation} \label{eqn:unwindingtr}
\wt{K}_{T}=(\psi_{T}+\eta_{T})^{-2}(\mu_{T}-\sigma_{T})+{\tau\over n}(\psi_{T}+\eta_{T})^{q}\gamma_{T}.
\end{equation}

Now if we let $s\in[-T/2,-1]$, the cutoffs $\chi_{j}\equiv\delta_{j,1}$, so $\gamma_{T}\equiv\gamma_{c}|_{B_{R}(p_{1})}$ and likewise $\mu_{T}\equiv\psi_{c}^{2}\mu|_{B_{R}(p_{1})}$. In all computations going forward, we regard $\gamma$, $h$, $\mu$, $\sigma_{T}$, $\eta_{T}$, and $\psi_{c}$ as being restricted to the image of $B_{R}(p_{1})$ under the gluing map. We also do the same for $r$, and use (\ref{eqn:psir1}) in the formula for $\psi_{T}$. We will also let $r_{0}$ be the parameter corresponding to $s_{0}$.

We make a small simplification before unwinding all the definitions: because $\wt{\gamma}_{T}$ is a conformal transformation of $\gamma$ outside $Q_{0}$, because the latter decomposes as a product between the radial and spherical components (\ref{eqn:normaldecomp}), and because $\wt{\mu}_{T}$ is traceless with respect to $\wt{\gamma}_{T}$, we have that
\begin{equation*}
(\wt{\gamma}_{T})^{ab}(\wt{\mu}_{T})_{ab}=-(\wt{\gamma}_{T})^{rr}(\wt{\mu}_{T})_{rr}.
\end{equation*}
Expanding everything out in terms of $r$ with (\ref{eqn:finalmetric}), (\ref{eqn:blownup}), (\ref{eqn:normaldecomp}), (\ref{eqn:psic}), (\ref{eqn:psir1}), and (\ref{eqn:eta}), we obtain:

\begin{align} \label{aln:trcomp}
\tr_{\wt{\gamma}_{T}|_{r=r_{0}}}\wt{K}_{T}&=(\wt{\gamma}_{T})^{ab}(\wt{K}_{T})_{ab} \\
&=-\wt{\psi}_{T}^{-2}(\wt{\gamma}_{T})^{rr}(\wt{\mu}_{T})_{rr}+(\wt{\gamma}_{T})^{ab}{\tau\over n}(\wt{\gamma}_{T})_{ab} \nonumber \\
&=-\wt{\psi}_{T}^{-2}\left({\psi_{c}\over\wt{\psi}_{T}}\right)^{q}\gamma^{rr}(\wt{\mu}_{T})_{rr}+{\tau(n-1)\over n} \nonumber \\
&=-\wt{\psi}_{T}^{{-2n\over n-2}}\psi_{c}^{q}(\wt{\mu}_{T})_{rr}+{\tau(n-1)\over n} \nonumber \\
&=-\wt{\psi}_{T}^{{-2n\over n-2}}r_{0}^{2}(\psi_{c}^{2}\mu-\sigma_{T})_{rr}+{\tau(n-1)\over n} \nonumber \\
&=-\wt{\psi}_{T}^{{-2n\over n-2}}r_{0}^{n}\mu_{rr}+\wt{\psi}_{T}^{{-2n\over n-2}}r_{0}^{2}(\sigma_{T})_{rr}+{\tau(n-1)\over n} \nonumber \\
&=-\left(r_{0}^{{n-2\over2}}+\wt{\chi}_{2}\left({R^{2}\over e^{T}r_{0}}\right)^{{n-2\over2}}+\eta_{T}\right)^{{-2n\over n-2}}r_{0}^{n}\mu_{rr} \nonumber \\
&\quad+\left(r_{0}^{{n-2\over2}}+\wt{\chi}_{2}\left({R^{2}\over e^{T}r_{0}}\right)^{{n-2\over2}}+\eta_{T}\right)^{{-2n\over n-2}}(\sigma_{T})_{tt}+{\tau(n-1)\over n}. \nonumber
\end{align}

A symmetric computation for $s\in[1,T/2]$ gives that

\begin{align*}
\tr_{\wt{\gamma}_{T}|_{r=r_{0}}}\wt{K}_{T}&=-\left(\wt{\chi}_{1}\left({R^{2}\over e^{T}r_{0}}\right)^{{n-2\over2}}+r_{0}^{{n-2\over2}}+\eta_{T}\right)^{{-2n\over n-2}}r_{0}^{n}\mu_{rr} \\
&\quad+\left(\wt{\chi}_{1}\left({R^{2}\over e^{T}r_{0}}\right)^{{n-2\over2}}+r_{0}^{{n-2\over2}}+\eta_{T}\right)^{{-2n\over n-2}}(\sigma_{T})_{tt}+{\tau(n-1)\over n},
\end{align*}
where now all objects are defined on $B_{R}(p_{2})\sm\{p_{2}\}$.

\subsubsection{Computing $H_{\wt{\gamma}_{T}|_{s=s_{0}}}$} \label{subsubsec:Hcomp}

We break this computation into two pieces: First we compute the mean curvature of a cross section just after gluing, and then we compute the final mean curvature after the second conformal transformation. In both computations, we use the following formula (from \cite{E92}) for mean curvature of the boundary of a manifold after a conformal transformation: If $\wt{g}=e^{2f}g$ is a conformally-related metric on a Riemannian manifold $(M,g)$ with boundary, then the resulting mean curvature is:
\begin{equation} \label{eqn:confmeancurv}
\wt{H}|_{\partial M}=e^{-f}\left(H|_{\partial M}+(n-1)\nu f\right),
\end{equation}
where $\nu$ is the outward-pointing unit normal to $\partial M$ with respect to $g$. As before, we restrict our computations to the portions of the tube where $\gamma_{T}$ is either $\gamma_{1}$ or $\gamma_{2}$. Again, we will focus on the left-hand side: that is, $\gamma_{T}\equiv\gamma_{1}=(\gamma_{c})|_{B_{R}(p_{1})\sm\{p_{1}\}}$.

From the decomposition of $\gamma$ in (\ref{eqn:normaldecomp}), we see that the unit normal pointing toward the center of the tube is $-\partial_{r}$. The first conformal transformation we apply is $e^{2f}=\psi_{c}^{-q}=r^{-2}$, so $f=-\log r$. This yields:
\begin{equation} \label{eqn:mcfirstconf}
H_{\gamma_{T}|_{r=r_{0}}}=r_{0}\left(H_{\gamma|_{r=r_{0}}}+{n-1\over r_{0}}\right)=r_{0}H_{\gamma|_{r=r_{0}}}+n-1.
\end{equation}

Focusing in on $H_{\gamma|_{r=r_{0}}}$, let $\{\partial_{1},\ldots,\partial_{n-1},\partial_{r}\}$ be the normal coordinate frame on $B_{R}(p_{1})\sm\{p_{1}\}$ used at the beginning of the gluing construction. Note that as above, we use beginning-alphabet letters for the cross-section $r=r_{0}$ (i.e. corresponding to $\{\partial_{1},\ldots,\partial_{n-1}\}$) and mid-alphabet letters for the entire frame. We obtain:
\begin{equation} \label{eqn:mcnabla}
H_{\gamma|_{r=r_{0}}}=\gamma^{ab}\langle\nabla_{\partial_{a}}-\partial_{r},\partial_{b}\rangle_{\gamma}=-\gamma^{ab}\langle\Gamma_{ar}^{c}\partial_{c},\partial_{b}\rangle_{\gamma}=-\gamma^{ab}\Gamma_{ar}^{c}\gamma_{cb}=-\Gamma_{ar}^{a}.
\end{equation}

This leaves us to compute $-\Gamma_{ar}^{a}$ in normal coordinates. Using the orthogonality properties of normal coordinates and the decomposition of $\gamma$, we obtain:
\begin{align*}
H_{\gamma|_{r=r_{0}}}=-\Gamma_{ar}^{a}&=-{1\over2}\gamma^{ai}\left(\partial_{a}\gamma_{ri}+\partial_{r}\gamma_{ai}-\partial_{i}\gamma_{ar}\right) \\
&=-{1\over2}\gamma^{ab}\left(\partial_{a}\gamma_{rb}+\partial_{r}\gamma_{ab}\right) \\
&=-{1\over2}\gamma^{ab}\left(\partial_{r}\gamma_{ab}\right) \\
&=-{1\over2r_{0}^{2}}h^{ab}(r_{0})\left(\partial_{r}(r^{2}h_{ab}(r))\right) \\
&=-{1\over2r_{0}^{2}}h^{ab}(r_{0})\left(2r_{0}h_{ab}(r_{0})+r_{0}^{2}h_{ab}'(r_{0})\right) \\
&={1-n\over r_{0}}-{1\over2}h^{ab}(r_{0})h_{ab}'(r_{0}).
\end{align*}
Plugging this into (\ref{eqn:mcfirstconf}), we get:
\begin{equation} \label{eqn:firstmc}
H_{\gamma_{T}|_{r=r_{0}}}=-r_{0}\left({n-1\over r_{0}}+{1\over2}h^{ab}(r_{0})h_{ab}'(r_{0})\right)+n-1=-{r_{0}\over2}h^{ab}(r_{0})h_{ab}'(r_{0}).
\end{equation}

This completes the computation of the mean curvature prior to the final conformal transformation. We remark that as $r_{0}\searrow0$, the mean curvatures of the cross-sections go to zero, since near the center of the gluing tube, the metric is very close to cylindrical. For the final transformation, we have $e^{2f}=(\psi_{T}+\eta_{T})^{q}$, so
\begin{equation} \label{eqn:finalconf}
f={q\over2}\log(\psi_{T}+\eta_{T})={2\over n-2}\log(\psi_{T}+\eta_{T}).
\end{equation}
In addition, the unit normal (pointing to the center of the tube) of $r=r_{0}$ with respect to $\gamma_{1}$ is $\partial_{t}=-r\partial_{r}$. Thus, by (\ref{eqn:confmeancurv}), we obtain
\begin{align} \label{aln:rawmc}
&H_{\wt{\gamma}_{T}|_{r=r_{0}}} \\
&=(\psi_{T}+\eta_{T})^{-{2\over n-2}}\left(H_{\gamma_{T}|_{r=r_{0}}}-(n-1)r_{0}\partial_{r}\left.\left({2\over n-2}\log(\psi_{T}+\eta_{T})\right)\right|_{r=r_{0}}\right) \nonumber \\
&=\left(r_{0}^{{n-2\over2}}+\wt{\chi}_{2}\left({R^{2}\over e^{T}r_{0}}\right)^{{n-2\over2}}+\eta_{T}\right)^{-{2\over n-2}} \nonumber \\
&\qquad\cdot\left(-{r_{0}\over2}h^{ab}(r_{0})h_{ab}'(r_{0})-{2(n-1)r_{0}\partial_{r}\left.\left(\log(\psi_{T}+\eta_{T})\right)\right|_{r=r_{0}}\over n-2}\right) \nonumber
\end{align}
We now investigate the $\partial_{r}(\log(\psi_{T}+\eta_{T}))$ term:
\begin{align*}
&\partial_{r}\left.(\log(\psi_{T}+\eta_{T}))\right|_{r=r_{0}} \\
&={\partial_{r}\left.\left(r^{{n-2\over2}}+\wt{\chi}_{2}\left({R^{2}\over e^{T}r}\right)^{{n-2\over2}}+\eta_{T}\right)\right|_{r=r_{0}}\over r_{0}^{{n-2\over2}}+\wt{\chi}_{2}\left({R^{2}\over e^{T}r_{0}}\right)^{{n-2\over2}}+\eta_{T}} \\
&={{n-2\over2}r_{0}^{{n-4\over2}}+\partial_{r}\wt{\chi}_{2}|_{r=r_{0}}\left({R^{2}\over e^{T}r_{0}}\right)^{{n-2\over2}}-\wt{\chi}_{2}\left({n-2\over2}\right)\left({R^{2}\over e^{T}}\right)^{{n-2\over2}}r_{0}^{-{n\over2}}+\partial_{r}\left.\eta_{T}\right|_{r=r_{0}} \over r_{0}^{{n-2\over2}}+\wt{\chi}_{2}\left({R^{2}\over e^{T}r_{0}}\right)^{{n-2\over2}}+\eta_{T}}.
\end{align*}
Thus, plugging the above into (\ref{aln:rawmc}) and using $(\partial_{r}\eta_{T})={\partial_{t}\eta_{T}\over r}$, our mean curvature is
\begin{align} \label{aln:mccomp}
&H_{\wt{\gamma}_{T}|_{r=r_{0}}} \\
&=\left(r_{0}^{{n-2\over2}}+\wt{\chi}_{2}\left({R^{2}\over e^{T}r_{0}}\right)^{{n-2\over2}}+\eta_{T}\right)^{{-2\over n-2}} \nonumber \\
&\cdot\left(\rule{0cm}{1.2cm}-{r_{0}\over2}h^{ab}(r_{0})h_{ab}'(r_{0})\right. \nonumber \\
&\left.\rule{0cm}{1.2cm}-{(n-1)\left(r_{0}^{{n-2\over2}}+\left({2\partial_{r}\wt{\chi}_{2}\over n-2}\right)\left({R^{2}\over e^{T}}\right)^{{n-2\over2}}r_{0}^{{4-n\over2}}-\wt{\chi}_{2}\left({R^{2}\over e^{T}r_{0}}\right)^{{n-2\over2}}-{2\partial_{t}\eta_{T}\over n-2}\right) \over r_{0}^{{n-2\over2}}+\wt{\chi}_{2}\left({R^{2}\over e^{T}r_{0}}\right)^{{n-2\over2}}+\eta_{T}}\right). \nonumber
\end{align}

Now for the right-hand side of the cylinder, the result is exactly the same, but with $\wt{\chi}_{1}$ in place of $\wt{\chi}_{2}$, and where $h$ and $\eta_{T}$ are computed on the right-hand side.

\subsubsection{Estimating the expansion} \label{subsubsec:estimate}

We now estimate the magnitudes of $\tr_{\wt{\gamma}_{T}|_{r=r_{0}}}\wt{K}_{T}$ and $H_{\wt{\gamma}_{T}|_{r=r_{0}}}$ for $r_{0}={R\over e^{T/2-1}}$ (that is, $s=-1$) to demonstrate how the null expansions close to the middle of the neck depend primarily on the geometry of the gluing neck. In all computations, we first write exactly what each term is, then estimate by throwing out terms that decay faster in $T$ than the others. In particular, we first estimate the $C^{k}$ norm of $\eta_{T}$ on the cross section $\{s=-1\}$ from the bound in Equation \ref{eqn:etabds}. First off, we have
\begin{equation*}
\left.||w_{T}^{-\delta}\eta_{T}||_{k}\right|_{\{s=-1\}}\le||w_{T}^{-\delta}\eta_{T}||_{k,\alpha}\lesssim e^{(-\lambda+\delta/q)T}
\end{equation*}
for fixed $\delta\in(0,1)$ and $\lambda>1/q$. Recalling from Definition \ref{def:wgtedHolder} that $w_{T}=e^{-T/q}\cosh(2s/q)$ on $C_{T}$, we see that on $\{s=-1\}$, the following bound holds on the weight:
\begin{equation*}
\left.||w_{T}^{-\delta}||_{k}\right|_{\{s=-1\}}\ge c(n,k,\delta)e^{\delta T/q}
\end{equation*}
This yields:
\begin{equation} \label{eqn:etacrossbd}
||\eta_{T}||_{k}|_{\{s=-1\}}\le C(n,k,\delta)e^{-\lambda T}.
\end{equation}

Now, when we choose $r_{0}={R\over e^{T/2-1}}$, we have $\wt{\chi}_{2}=1$ and $\partial_{r}\wt{\chi}_{2}=0$. We also note since we can rewrite
\begin{equation*}
{R^{2}\over e^{T}\left({R\over e^{{T\over2}-1}}\right)}={R\over e^{{T\over2}+1}},
\end{equation*}
we can isolate the dependence of $\psi_{T}(r_{0})$ on $T$ as follows:
\begin{align} \label{aln:psiTr0}
\psi_{T}(r_{0})&=\left({R\over e^{{T\over2}-1}}\right)^{{n-2\over2}}+\left({R\over e^{{T\over2}+1}}\right)^{{n-2\over2}} \\
&=\left((Re)^{{n-2\over2}}+\left({R\over e}\right)^{{n-2\over2}}\right)e^{-{T\over q}}. \nonumber
\end{align}
Thus, comparing (\ref{eqn:etacrossbd}) with (\ref{aln:psiTr0}), we conclude that the contribution of $\eta_{T}$ to $\wt{\psi}_{T}$ is negligible on the cross-section $\{s=-1\}$.

Now for the $\tr\wt{K}_{T}$ term from (\ref{aln:trcomp}), using (\ref{eqn:sigmaHolderbd}) to estimate $\sigma_{T}$, we have:
\begin{align} \label{aln:trKbd}
&\left|\tr_{\wt{\gamma}|_{r={R\over e^{{T\over2}-1}}}}\wt{K}_{T}\right| \\
&=\left|-\left(  \left((Re)^{{n-2\over2}}+\left({R\over e}\right)^{{n-2\over2}}\right)e^{-{(n-2)T\over 4}}   +\eta_{T}\right)^{{-2n\over n-2}}\left({R\over e^{{T\over2}-1}}\right)^{n}\mu_{rr}\right. \nonumber \\
&\quad\;\;+ \left. \left(  \left((Re)^{{n-2\over2}}+\left({R\over e}\right)^{{n-2\over2}}\right)e^{-{(n-2)T\over 4}}   +\eta_{T}\right)^{{-2n\over n-2}}   (\sigma_{T})_{tt}+{\tau(n-1)\over n}\right| \nonumber \\
&\le C_{1}(R,n,k,\delta,\mu)\left(e^{-{(n-2)T\over4}}\right)^{{-2n\over n-2}}\left(e^{-{T\over2}}\right)^{n} \nonumber \\
&\qquad+C_{2}(R,n,k,\delta)\left(e^{-{(n-2)T\over4}}\right)^{{-2n\over n-2}}T^{3}e^{{-nT\over2}}+{\tau(n-1)\over n} \nonumber \\
&=C_{1}(R,n,k,\delta,n)+C_{2}(R,n,k,\delta)T^{3}+{\tau(n-1)\over n} \nonumber \\
&\lesssim T^{3}, \nonumber
\end{align}
where $C_{1}$ and $C_{2}$ are positive constants.

Now we move to the computation of mean curvature at $r_{0}={R\over e^{{T\over2}-1}}$:
\begin{align} \label{aln:mcbd}
&\left|H_{\wt{\gamma}_{T}|_{r={R\over e^{{T\over2}-1}}}}\right| \\
&=\left|\rule{0cm}{1.2cm}\left(\left((Re)^{{n-2\over2}}+\left({R\over e}\right)^{{n-2\over2}}\right)e^{-{(n-2)T\over4}}+\eta_{T}\right)^{{-2\over n-2}}\right. \nonumber \\
&\qquad\cdot\left(\rule{0cm}{1.2cm} -\left({R\over2e^{{T\over2}-1}}\right)h^{ab}\left({R\over e^{{T\over2}-1}}\right)h_{ab}'\left({R\over e^{{T\over2}-1}}\right)      \right. \nonumber \\
&\left.\left.\rule{0cm}{1.2cm}\qquad\qquad   -{(n-1)\left(  \left((Re)^{{n-2\over2}}-\left({R\over e}\right)^{{n-2\over2}}\right)e^{-{(n-2)T\over4}}-{2\partial_{t}\eta_{T}\over n-2}\right)\over\left((Re)^{{n-2\over2}}+\left({R\over e}\right)^{{n-2\over2}}\right)e^{-{(n-2)T\over4}} +\eta_{T}}\right)\right| \nonumber
\end{align}
The above expression is of the form $|X_{1}(T)(X_{2}(T)+X_{3}(T))|$. To estimate it, we consider each piece separately. First off, we see that
\begin{equation} \label{eqn:X1bd}
X_{1}\ge c(R,n,k,\delta)\left(e^{-{(n-2)T\over4}}\right)^{{-2\over n-2}}=c(R,n,k,\delta)e^{T/2},
\end{equation}
where $c$ is a positive constant. Now, while we ultimately seek a lower bound on the mean curvature, we aim to show that $X_{3}$ dominates $X_{2}$, so consider the following upper bound on $|X_{2}|$:
\begin{equation} \label{eqn:X2bd}
|X_{2}|\le C(R,n,h,h')e^{-T/2}.
\end{equation}
We remark that in fact, $h'\to0$ as $r\to0$, so we would not be able to bound $|X_{2}|$ below by $ce^{-T/2}$. Now we move to $|X_{3}|$: since the numerator term is bounded below by $c(R,n,k,\delta)e^{-T/q}$ and the denominator is bounded above by $C(R,n,k,\delta)e^{-T/q}$, we observe
\begin{equation} \label{eqn:X3bd}
|X_{3}|\ge c(R,n,k,\delta)
\end{equation}
for some positive constant $c$. Combining (\ref{aln:mcbd}), (\ref{eqn:X1bd}), (\ref{eqn:X2bd}), and (\ref{eqn:X3bd}), we obtain:
\begin{equation} \label{eqn:mcbd}
\left|H_{\wt{\gamma}_{T}|_{r={R\over e^{{T\over2}-1}}}}\right|\gtrsim e^{T/2}
\end{equation}

Thus, since $e^{{T\over2}}$ dominates $T^{3}$ for $T$ large, the null expansion,
\begin{equation*}
\theta_{s=-1}=\tr_{\wt{\gamma}_{T}|_{s=-1}}\wtK_{T}+H_{\wt{\gamma}_{T}|_{s=-1}},
\end{equation*}
at the cross section $s=-1$ is dominated by the mean curvature term for large enough gluing parameter. Note also that in the above computation of the mean curvature (relative to the unit normal pointing to the center of the neck), the terms that dominate do not have signs depending on the local geometry: we in fact have that
\begin{equation*}
H_{\wt{\gamma}_{T}|_{s=-1}}<-Ce^{{T\over2}},
\end{equation*}
where $C$ is a positive constant. Likewise, we have the symmetric estimate:
\begin{equation*}
H_{\wt{\gamma}_{T}|_{s=1}}<-\wt{C}e^{{T\over2}},
\end{equation*}
where $\wt{C}$ is again a positive constant. This yields that $\theta_{s=-1}$ and $\theta_{s=1}$ are both outer trapped, as desired.

\subsection{Inferring future and past null incompleteness} \label{subsec:incompleteness}

Passing to a noncompact cover $\pi:(\cal{N},\pi^{*}\wt{\gamma}_{T},\pi^{*}\wt{K}_{T})\to(\wtM,\wt{\gamma}_{T},\wtK_{T})$, we may consider the corresponding outer trapped surfaces in $\cal{N}$ and use Theorem \ref{thm:outerpenrose} to deduce future null incompleteness of any spacetime development $\wt{\cal{N}}$. In addition, applying a time reversal, we may also conclude past null incompleteness.

It remains to show that any spacetime evolution $\wh{M}$ of $\wtM$ is future null geodesically incomplete. We use the following lemma from \cite{GL18(2)}:
\begin{lemma}
Let $(\wtM,\wt{\gamma},\wt{K})$ be a smooth spacelike Cauchy surface in a spacetime $(\wh{M},g)$, and suppose $\pi:\cal{N}\to \wtM$ is a Riemannian covering map. Then there exists a Lorentzian covering map $\wh{\pi}:\wh{\cal{N}}\to\wh{M}$ extending $\pi$ such that $(\cal{N},\pi^{*}\wt{\gamma},\pi^{*}\wt{K})$ is a Cauchy surface for the spacetime $\wh{\cal{N}}$.
\end{lemma}

\begin{center}
\begin{tikzcd}[ampersand replacement=\&]
(\cal{N},\pi^{*}\wt{\gamma},\pi^{*}\wt{K})\arrow{d}{\pi} \arrow[hook]{r} \& (\wh{\cal{N}},\wh{\pi}^{*}g)\arrow{d}{\wh{\pi}} \\
(\wtM,\wt{\gamma},\wt{K})\arrow[hook]{r} \& (\wh{M},g)
\end{tikzcd}
\end{center}

By the same geodesic lifting argument as posed in \cite{BLP19}, we conclude that future and past null incompleteness of $\wh{\cal{N}}$ implies future and past null incompleteness of $\wh{M}$, as desired.

%Now suppose $\wh{M}$ is future null geodesically complete. Since $\wh{\cal{N}}$ is future null geodesically incomplete, there exists a future inextendible smooth null geodesic $\zeta:[0,\alpha)\to\wh{\cal{N}}$ that terminates at affine parameter $\alpha<\infty$. Consider the smooth null geodesic $\wh{\pi}(\zeta)\subset\wh{M}$. Then by future null completeness of $\wh{M}$, we must be able to find a smooth null geodesic $\wh{\zeta}:[0,\infty)\to\wh{M}$ extending $\wh{\pi}(\zeta)$. Let $\veps>0$ be small enough so that $\wh{\zeta}(\alpha-\veps,\alpha+\veps)$ is contained in a single evenly-covered neighborhood $U\subset\wh{M}$, and pick a smooth local section $\sigma:U\to\wh{\cal{N}}$ of the covering such that $\zeta(t)=\sigma\circ\wh{\pi}\circ\zeta(t)$ for every $t\in(\alpha-\veps,\alpha)$. Then we see that the smooth null geodesic $\xi:[0,\alpha+\veps)\to\wh{\cal{N}}$ defined by
%\begin{equation*}
%\xi(t):=\begin{cases}
%\zeta(t) & t\in[0,\alpha) \\
%\sigma\circ\wh{\zeta}(t) & t\in(\alpha-\veps,\alpha+\veps)
%\end{cases}
%\end{equation*}
%extends $\zeta$ to the future, which contradicts our assumption on $\zeta$. Thus, we must indeed have that any spacetime evolution of $(\wtM,\wt{\gamma},\wtK)$ is future null geodesically incomplete.

\subsection{Inferring a stable MOTS} \label{subsec:MOTSexist}

In addition to incompleteness, we may now use our computations of the null expansion at $s=\pm1$ to apply the MOTS existence result of \cite{AEM11}. Indeed, the subset $Q_{0}\subset\wt{M}$ in the center of the neck has boundary consisting of the cross sections $\{s=\pm1\}$, and when we consider the mean curvature of these cross sections with respect to the unit normal pointing out of $Q_{0}$ (denoted $\wh{H}=-H$, where $H$ is as above), we have
\begin{equation*}
\wh{H}_{\wt{\gamma}_{T}|_{s=-1}}-\tr_{\wt{\gamma}_{T}|_{s=-1}}\wt{K}_{T}>0\quad\text{and}\quad \wh{H}_{\wt{\gamma}_{T}|_{s=1}}+\tr_{\wt{\gamma}_{T}|_{s=1}}\wt{K}_{T}>0.
\end{equation*}
Thus, by Theorem 3.3 of \cite{AEM11}, there exists a $C$-almost minimizing boundary (for $C=C(|\wt{K}_{T}|_{C(\ol{Q_{0}})}$) $\Sigma^{n-1}\subset Q_{0}$ with a singular set of Hausdorff codimension at most 7, satisfying $H_{\Sigma}+\tr_{\Sigma}\wt{K}_{T}=0$ (calculated with respect to the unit normal pointing toward $\{s=1\}$) distributionally. For $3\le n\le7$, $\Sigma$ is a smooth closed embedded hypersurface homologous to $\{s=1\}$ and a stable MOTS.

\section{Open questions} \label{sec:fut}

The following are open questions related to the results obtained thus far:

\begin{itemize}
\item As mentioned in the Introduction, there is still work to do to obtain null incompleteness in the general, non-vacuum gluing constructions.

\item For incompleteness to follow from existence of an outer trapped surface, we require both (1) that the glued Cauchy surface $\wtM$ is either noncompact itself or has nontrivial fundamental group, and (2) that $\wtM=M_{1}\#M_{2}$ so that we can (1) pass to a noncompact covering space and (2) satisfy the `separating' restriction of Theorem \ref{thm:outerpenrose}. Heuristically, however, the geometry is so warped around the gluing neck that even without a noncompact cover or the separating condition, we should expect null incompleteness. It would be desirable to obtain a new incompleteness theorem that encompasses these cases.

\item In order for \cite{BLP19} to provide supportive evidence for the Bartnik splitting conjecture, we would wish to obtain that those spacetimes are timelike incomplete as well as null incomplete. This seems to be a much more difficult endeavor from the initial data perspective of the gluing constructions.
\end{itemize}

\end{document}